\documentclass[slac_one]{revtex4}
\usepackage{graphicx}
\usepackage{fancyhdr}
\begin{document}

\title{Study of Rare Exclusive Electroweak Processes at HERA} 

%

\author{Z. Zhang (for the H1 and ZEUS Collaborations)}
\affiliation{LAL, Universit\'e Paris-Sud et IN2P3/CNRS, 91898 Orsay Cedex, FRANCE}
%

\begin{abstract}
Results on multi-lepton events at high transverse momenta, 
isolated lepton events with large missing transverse energy and 
single W production  reported to this conference are based for the first time 
on the full data samples taken by two colliding experiments, H1 and ZEUS,
at HERA. 
The data correspond to an integrated luminosity of about $1\,{\rm fb}^{-1}$
from both experiments.
\end{abstract}

\maketitle

\thispagestyle{fancy}


\section{INTRODUCTION} 

The data taking at HERA, where electrons or positrons of $27.5\,{\rm GeV}$ 
collided with protons of up to $920\,{\rm GeV}$, ended in June 2007. 
Each of the H1 and ZEUS experiments collected around $0.5\,{\rm fb}^{-1}$ 
data from 
the whole running period 1992-2007. The largest samples are from the second 
data taking period 2003-2007 (HERA-2). In comparison with HERA-1, 
the integrated luminosity of the $e^+p$ 
and $e^-p$ samples has a 2- and 10-fold increase, respectively. In addition, 
the $e^+$ and $e^-$ beams at HERA-2 were longitudinally polarised. 
These data samples have made possible both the study of rare exclusive 
electroweak processes with cross section values down to $1\,{\rm pb}$ and 
the search for new physics phenomena.

This talk covers eight abstracts submitted to this conference from H1 and ZEUS
on three main topics listed in the abstract. The results presented at 
the conference are briefly summarised here in the following sections.

\section{MULTI-LEPTON EVENTS AT HIGH TRANSVERSE MOMENTA}

An excess of multi-lepton events at high $P_T$ at HERA was first reported 
in~\cite{h1-epjc31} by H1 based on HERA-1 data. The dominant Standard Model 
(SM) processes are from the lepton pair production in photon-photon 
interactions, $\gamma\gamma\rightarrow l^+l^-$, where the photons are 
radiated from incident beam particles. The background contributions are 
mainly from neutral current deep inelastic scattering (DIS) and QED 
compton processes where in addition to genuine electrons, hadrons or radiated 
photons are misidentified as electrons or muons. Beyond the SM, the production 
of a doubly charged Higgs boson~\cite{ap94-h1-plb638} or processes involving 
generic bosons carrying two units of lepton number (bi-leptons)~\cite{cd98} 
could lead to multi-leptons events of large invariant mass.

The analyses are performed in a model independent way with the following main 
selection cuts. Take H1~\cite{h1-plb} as an example, each event has to have 
at least two central ($20^o<\theta <150^o$) electron or muon candidates with 
the leading lepton $P^1_T>10\,{\rm GeV}$, the other lepton 
$P^2_T>5\,{\rm GeV}$ and additional electrons in an extended angular region 
$5^o<\theta <175^o$ and 
additional muons in $20^o<\theta <160^o$ and $P_T>2\,{\rm GeV}$.
  
H1 has analysed seven topologies in $ee$, $\mu\mu$, $e\mu$, $eee$, $e\mu\mu$,
$ee\mu$ and $eeee$. In all the topologies, the observed event yields are 
found in good agreement with the predicted ones~\cite{h1-plb}. 
However, when the comparison is made for the invariant mass of two highest 
$P_T$ leptons $M_{12}>100\,{\rm GeV}$, excesses are observed in most of 
the topologies (Table~\ref{table1}) although the number of observed events 
remains statistically limited. Also shown in Table~\ref{table1} are 
preliminary results from ZEUS on di-electron and tri-electron 
samples~\cite{zeus-prel-07-022}. In both samples no excess has been observed.

\begin{figure}
\begin{tabular}{cc}
\includegraphics[width=70mm]{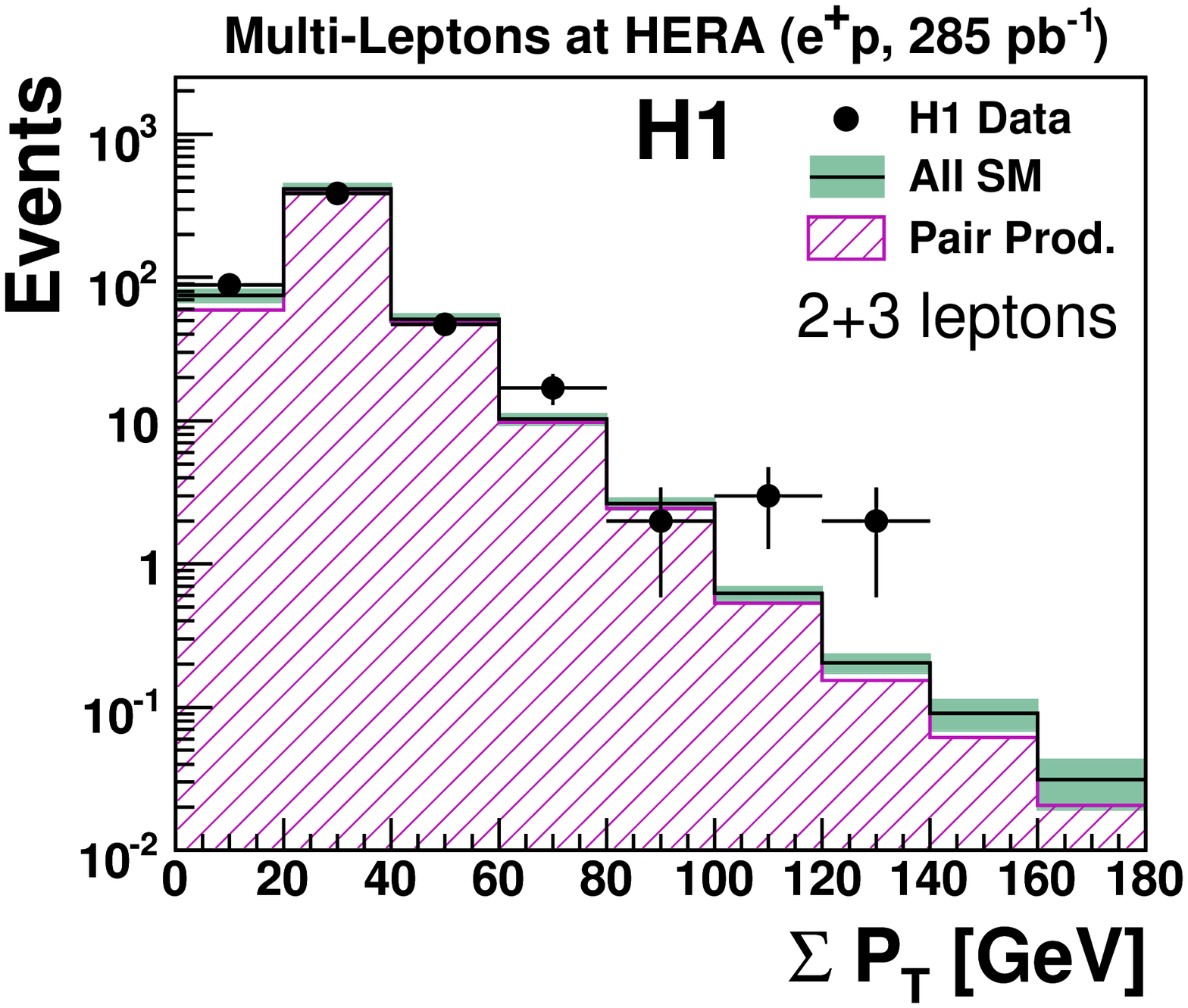} &
\includegraphics[width=70mm]{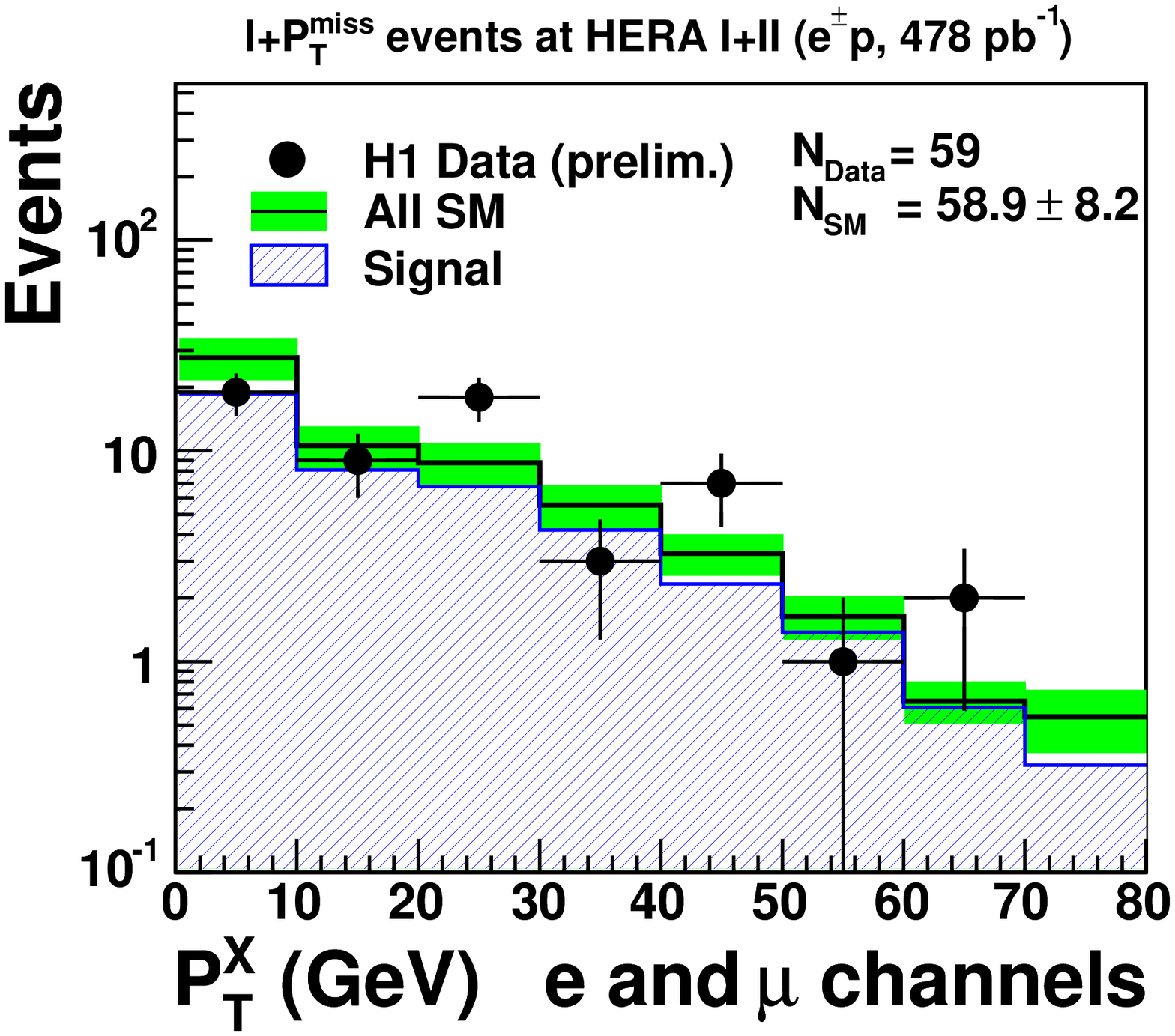}
\end{tabular}
\caption{\label{h1}Left: The distribution of the scalar sum of the transverse 
momenta $\sum P_T$ for di-lepton and tri-lepton event samples for $e^+p$ is
compared to the total SM expectation and the dominant pair production.
Right: The hadronic transverse momentum distribution in the isolated electron 
and muon channels combined is compared to the total SM expectation and
dominant real $W$ production.}
\end{figure}

H1 has also compared the distributions of the scalar sum of the transverse 
momentum ($\sum P_T$) (see e.g. Fig.~\ref{h1}(left) for the $e^+p$ data). 
At $\sum P_T>100\,{\rm GeV}$, 5 events have been 
observed in the $e^+p$ sample with 0$.96\pm 0.12$ expected. 
None has been observed in the $e^-p$ sample, however, while $0.64\pm 0.09$ 
events are expected. Therefore the excess is only shown in the $e^+p$ data 
sample.

Differential cross sections as a function of the leading transverse momentum 
$P^1_T$ for electron and muon pair production are measured by H1~\cite{h1-plb} 
in a restricted phase space dominated by photon-photon interactions 
($P^1_T>10\,{\rm GeV}$, $P^2_T>5\,{\rm GeV}$, $20^o<\theta <150^o$ , 
the inelasticity variable $y<0.82$ and the four-momentum transfer squared 
$Q^2<1\,{\rm GeV}^2$). ZEUS has released their preliminary 
results~\cite{zeus-prel-08-006} for this conference in di-muon channel with 
a slightly different phase space cut ($P_T>5\,{\rm GeV}$). Both H1 and ZEUS 
measure steeply falling cross sections in good agreement with the SM 
expectations.

\begin{table}[t]
\begin{center}
\caption{The number of observed events and SM expectations in different 
multi-lepton topologies for $M_{12}>100\,{\rm GeV}$. The numbers shown in 
parentheses correspond to the contribution from the dominant pair production 
in $\gamma\gamma$
interactions.}
\begin{tabular}{|c|c|c|c|c|}
\hline \textbf{Topology} & \multicolumn{2}{|c|}{\textbf{H1}} & \multicolumn{2}{|c|}{\textbf{ZEUS}} \\ \cline{2-5}
 & \textbf{Data} & \textbf{SM (pair)} & \textbf{Data} & \textbf{SM (pair)} \\
\hline
$ee$ & $3$ & $1.34\pm 0.20 (0.83)$ & $2$ & $1.7\pm 0.2 (0.9)$ \\ \hline
$e\mu$ & $1$ & $0.59\pm 0.06 (0.59)$ & & \\ \hline
$eee$ & $3$ & $0.66\pm 0.09 (0.66)$ & $2$ & $1.0\pm 0.1 (1.0)$ \\ \hline
$\mu\mu$ & $1$ & $0.17\pm 0.07 (0.17)$ & & \\ \hline
$e\mu\mu$ & $2$ & $0.16\pm 0.20 (0.83)$ & & \\ \hline
\end{tabular}
\label{table1}
\end{center}
\end{table}

\section{ISOLATED LEPTON EVENTS WITH LARGE MISSING ENERGY}

The first isolated muon event with large missing transverse energy was 
reported by H1 in 1994~\cite{h1-94}. Since then several searches have been 
performed by H1 and ZEUS for different types of isolated lepton ($e$, $\mu$,
$\tau$) and with increasingly large data samples~\cite{h1-zeus}. 
The results~\cite{h1prel,zeus08} reported to this conference with the full 
HERA data are summarised in Table~\ref{table2}. The H1 excess in the $e$ and
$\mu$ channels (Fig.~\ref{h1}(right)) is not confirmed by ZEUS. 
ZEUS has, however, observed $2$ 
isolated $\tau$ lepton events with a hadronic final state at large transverse 
momentum ($P^X_T>25\,{\rm GeV}$) for $0.20\pm0.05$ expected based on the 
HERA-1 data, of which $49\%$ is contributed by the SM signal events 
from single $W$ 
production with genuine isolated $\tau$ and missing transverse momentum in 
the final state. This number reflects thus the purity of the selection.
The corresponding signal contribution (purity) for the $e$ and $\mu$ channels 
is also shown in Table~\ref{table2}. 
The rest is considered as background due to 
misidentification or mismeasurement. 
This latter contribution includes neutral and charged 
current DIS events, lepton pair production and photoproduction of jets. 

The study of isolated lepton events with large missing transverse energy
is interesting as it allows constraints to be placed on
anomalous single top production~\cite{1top} and bosonic stop decays~\cite{stop}
at HERA.

\begin{table}[t]
\begin{center}
\caption{The number of observed events and expected contributions for
three types of isolated lepton in $e^\pm p$ data samples for $P^X_T>25\,{\rm GeV}$. The SM signal (dominated by $W$ production) is shown in percentage
in parentheses.}
\begin{tabular}{|c|c|c|c|c|c|}
\hline \textbf{Dataset} & \textbf{Lepton} & \multicolumn{2}{|c|}{\textbf{H1~\cite{h1prel}}} & \multicolumn{2}{|c|}{\textbf{ZEUS~\cite{zeus08}}} \\ \cline{3-6}
 & & \textbf{Data} & \textbf{Exp (signal)} & \textbf{Data} & \textbf{Exp (signal)} \\ \hline
 & $e$ & $11$ & $4.7\pm 0.9 (75\%)$ & $3$ & $4.0\pm 0.6 (77\%)$ \\ \cline{2-6}
$e^+p$ & $\mu$ & $10$ & $4.2\pm 0.7 (85\%)$ & $3$ & $3.4\pm 0.5 (81\%)$ \\
\cline{2-6}
 & $\tau$ & $0$ & $0.5\pm 0.1 (72\%)$ & & \\ \hline
 & $e$ & $3$ & $3.8\pm 0.6 (61\%)$ & $3$ & $3.2\pm 0.5 (69\%)$ \\ \cline{2-6}
$e^-p$ & $\mu$ & $0$ & $3.1\pm 0.5 (74\%)$ & $2$ & $2.3\pm 0.4 (85\%)$ \\
\cline{2-6}
 & $\tau$ & $1$ & $1.0\pm 0.1 (63\%)$ & & \\ \hline
\end{tabular}
\label{table2}
\end{center}
\end{table}

\section{W BOSON PRODUCTION CROSS SECTION}

Within the SM, events with an isolated electron or muon and large missing 
transverse energy arise dominantly from the real $W$ production. 
The $W$ production cross section can thus be determined. The results
from ZEUS~\cite{zeus08} and H1~\cite{h1prel-w} are
$\sigma(ep\rightarrow lWX)=(0.89^{+0.25}_{-0.22}\pm 0.10)\, {\rm pb}$ and
$\sigma(ep\rightarrow lWX)=(1.23\pm 0.25\pm 0.22)\,{\rm pb}$ quoted
at a center-of-mass energy of $316\,{\rm GeV}$ and 
$320\,{\rm GeV}$, respectively.
The first error is statistical and the second systematic.
Both measurements reach a significance of about $5\sigma$. 
They are in good agreement with the corresponding SM expectations of
$1.2\,{\rm pb}$ and $1.3\,{\rm pb}$ with an uncertainty of $15\%$ from
the next-to-leading order calculations.

To further test the compatibility of the observed $W$ decays with the SM,
a measurement of the $W$ boson polarisation is performed for the first
time by H1~\cite{h1prel-w}. The measurement makes use of the angular 
distribution of the $W$ boson production. For $W^+$ bosons, this reads:
\begin{equation}\label{eq:units}
\frac{dN}{d\cos\theta^\ast}\propto F_+\cdot\frac{3}{8}(1+\cos\theta^\ast)^2+F_0\cdot\frac{3}{4}(1-\cos^2\theta^\ast)+F_-\cdot\frac{3}{8}(1-\cos\theta^\ast)^2 \nonumber
\end{equation}
where $\theta^\ast$ is defined as the angle between the $W$ boson momentum
in the lab frame and that of the charged decay lepton in the $W$ boson rest
frame, $F_+$, $F_0$ and $F_-$ stand respectively for the right-handed 
polarisation fraction, the longitudinal and left-handed one with
$F_++F_0+F_-=1$. For $W^-$ bosons, the $\cos\theta^\ast$ distributions have
opposite values.

\begin{figure}
\begin{tabular}{cc}
\includegraphics[width=70mm]{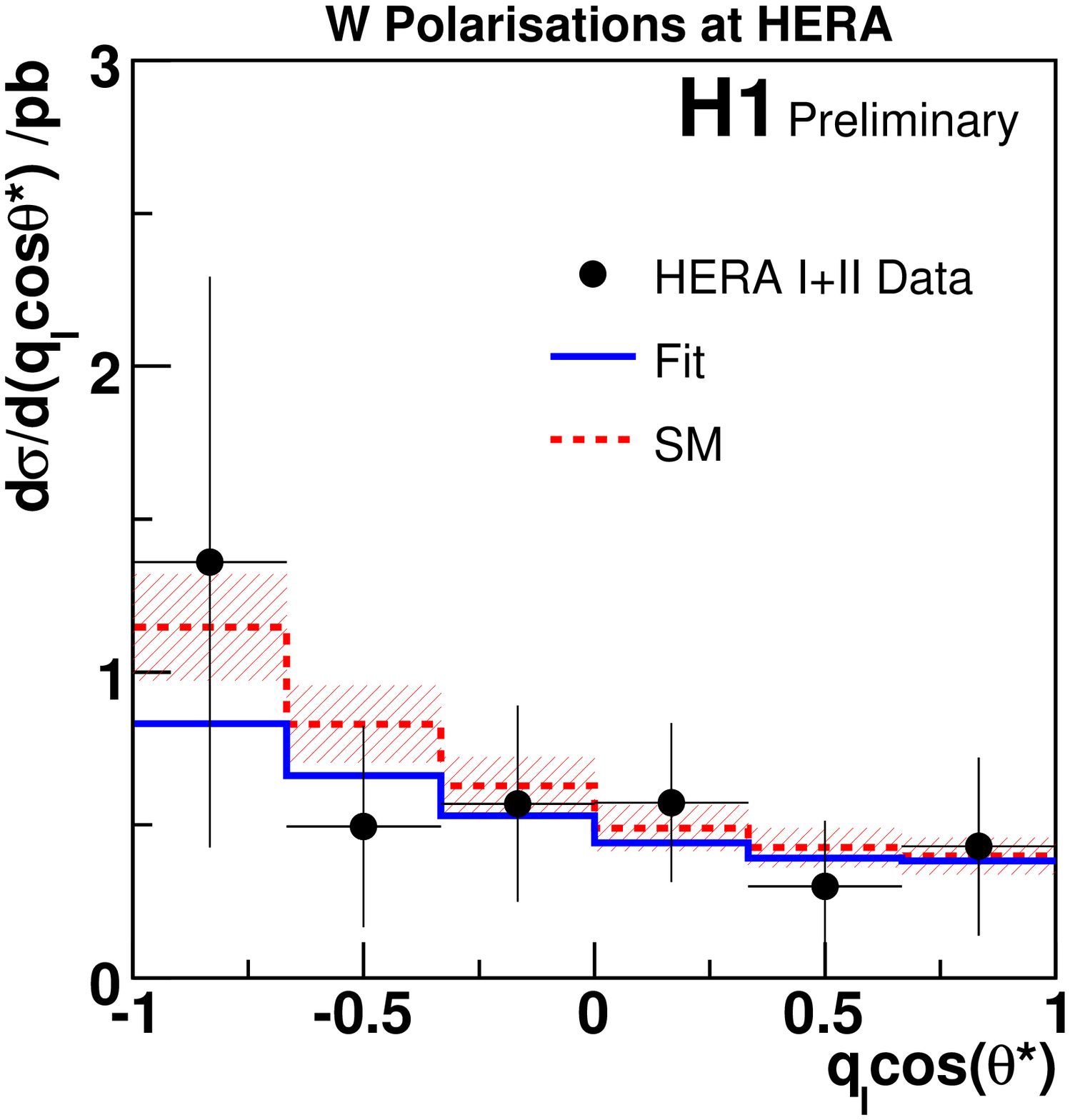} &
\includegraphics[width=70mm]{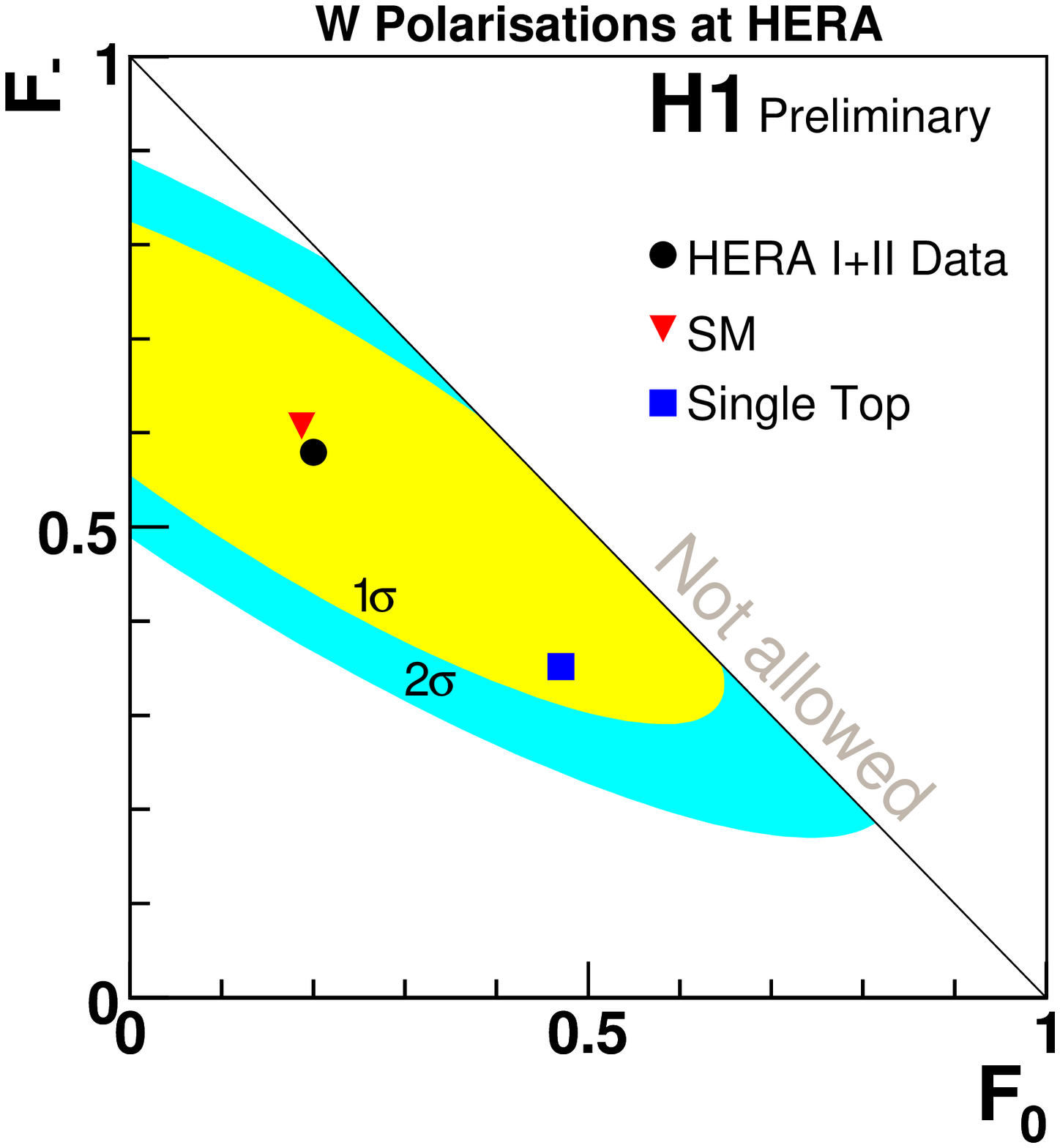}
\end{tabular}
\caption{\label{w}Left: Measured differential cross section shown with the 
statistical error bars in comparison with the SM prediction and the
result of the fit. Right: The fitted $F_-$ and $F_0$ at $1$ and $2\sigma$
contours in comparison with the SM prediction and anomalous single top
production.}
\end{figure}

The measured differential cross section is compared with the SM
expectation in Fig.~\ref{w}(left).
A fit to the distribution allows $F_-$ and $F_0$ being simultaneously 
extracted.
The results are shown in Fig.~\ref{w}(right) and found to be in good agreement
with the SM and compatible with anomalous single top production via 
flavour changing neutral current.

\section{SUMMARY}

Previously observed excesses in multi-lepton events at high transverse momenta
and isolated lepton events with large missing transverse energy by H1 
remain true with the full HERA data sample. The largest excess is up to 
about $3$ standard deviations and is however not confirmed by ZEUS.
Attempts in combining the H1 and ZEUS data have started and are being 
pursued~\cite{web}. As the HERA data
taking has ended, it is unlikely that a definitive conclusion can be drawn
with the combined data.
Future experiments will eventually tell us whether the excess is a purely
statistical fluctuation or a first sign of new physics.

\end{document}